# A Nonanticipative Analog Method for Long-Term Forecasting of Air Temperature Extremes

Dmytro Zubov[1]

Humberto A. Barbosa[2]

Gregory S. Duane[3,4]

[1] University of Information Science and Technology "St. Paul the Apostle", Republic of Macedonia; email: dzubov@ieee.org

[2] Laboratório de Análise e Processamento de Imagens de Satélites (LAPIS), Instituto de Ciências Atmosféricas, Universidade Federal de Alagoas, Brasil; email: barbosa33@gmail.com

[3] University of Colorado, Boulder, CO, USA; email: gregory.duane@colorado.edu

[4] Geofysisk Institutt, Universitetet i Bergen, Norway

**Abstract.** A nonanticipative analog method is used for the long-term forecast of air temperature extremes. The data to be used for prediction include average daily air temperature, mean visibility, mean wind speed, mean dew point, maximum and minimum temperatures reported during the day from 66 places around the world, as well as sea level, average monthly Darwin and Tahiti sea level pressures, SOI, equatorial SOI, sea surface temperature, and multivariate ENSO index. Every dataset is split into two samples – learning (1973-2010) and validation (2011-2013). Initially, the sum of variables in datasets for two locations, minus corresponding climatological values, is calculated over a summation interval of length from 1 to 365 days. A "quality criterion" selects datasets for two locations with appropriate lead-time and summation interval, which have maximum (or minimum) sum compared with the rest of data four times at least, when extreme events occur later within the learning sample. Up to 18.2% of all extremes are specifically predicted. The methodology has 100% accuracy with respect to the sign of predicted and actual values. It is more useful than current methods for predicting extreme values because it does not require the estimation of a probability distribution from scarce observations.

## 1. Introduction

Securing society against disasters is one of the central functions of the noosphere. There is barely any societal sector which is not to some extent concerned by extreme air temperatures and their long-term forecast together with related resilience and security issues. Long-term forecast models give an understanding of dependencies among different remote places and variables which are measured with significant lead-time (Zubov 2013). In addition, they allow prediction of possible natural disasters (e.g. Rocheva (2012)) and the taking of appropriate preventive measures as necessary. Air temperature has a great influence on the load of power service (Robinson 1997), and predictions are used for estimating future fuel needs. More critically, heat waves may produce significant disruptions to agricultural industries (Hudson et al. 2011) and provoke heart problems (peaks of cardiac diseases). Thus, the main objective of this work is to improve extreme air temperature long-term forecasting methodology using a nonanticipative approach (Greengard and Ruszczynski 2002), i.e. one in which no new data is assimilated between the last observations and the distant verification time. The loss of human life, as well as environmental, economic and material damage from extreme air temperatures could then be reduced. Even a small success rate implies a large socio-economic benefit.

Nowadays, a large set meteorological variables (air temperature, precipitation, wind, pressure, visibility, snow depth, etc. at different locations) is used for forecasting (Kattsov 2010). They interact constantly, and some variables may be evaluated using the others, in accordance with known teleconnection patterns (e.g. Nada Pavlovic Berdon (2013)). Thus, the reasoning of forecast models must involve the full set of meteorological variables. However, temperature and precipitation are the targets of long-term forecasting, mainly because of practical needs. Precipitation has a close relationship to air temperature and vice versa (Van Den Dool and Nap 1985). Correlation analysis shows that precipitation forecasting is effective within two weeks, air temperature over a much longer period (Zubov and Vlasov 2004). The impact is increased further because extremes can be used for the correction of forecasted averages.

A wide spectrum of forecast models has now been developed (Vilfand et al. 2003). They are usually classified into synoptic (e.g. Vorobiov (1991)), hydrodynamic (e.g. Belov et al. (1989)), and statistical (e.g. Onwubolu (2007)) groups. The first two are used only for short and medium-term forecasting mainly because they produce significant errors at long term (more than 20 % of the mean) and use highly complex equations. Heterogeneous algorithms are used for long-term weather forecasting – seasonal time series (Qiang Song 2011), neural networks (Gyanesh Shrivastava et al. 2012), probability theory (Sadokov et al. 2011), ensemble forecasting (Astahova and Alferov 2008), distinct scenarios of anthropogenic forcing (Bardin 2011), and dependency on the ENSO cycle (Higgins et al. 2004), etc. The nonlinearity and sensitivity of existing forecast models, possible small errors in initial conditions (dust, sand, pollution), random observation errors, background states, and lack of data combine to reduce the forecast accuracy and complicate the design of models (Douglas and Englehart 2007; Fathalla A Rihan and Chris G Collier 2010; Tyndall et al. 2010).

Inductive modelling shows good results when enough of the data that would be needed in more conservative approaches is fundamentally not obtainable. Self-organizing systems based on inductive modeling have already been applied to long-range weather forecasting (Madala and Ivakhnenko 1994). Successful applications of inductive modeling have been realized in other fields as well (e.g. stock market, economic systems, noise immunity, decision trees, data mining and neural networks). In (Zubov 2013), it was shown that robust highly accurate long-term forecasting of average daily air temperatures might be achieved using inductive modelling. The principle used in that work to predict high-impact weather events substantiates the interaction of different climate system components centered in different places. The first stage of the forecast model reasoning is the selection of three most data-related places using the Pearson product-moment correlation coefficient, which has to be greater than 0.8 in absolute value. The second stage is finding weighting coefficients of the forecast model to use with the inductive modelling criterion "minimum of regularity plus maximum of conjunctions" with a combinatorial algorithm. This approach corresponds to the phenomenon of teleconnections (Glantz 1991) because of a linkage between weather changes occurring in widely separated regions of the globe, even where a theoretical explanation for those linkages is not yet available.

Analog methods, in which a forecasted event is expected to mimic a recent event, have been applied for medium- and long-term forecasts. In (Toth 1989), an analog selection method relying on the coincidence of main features (large-scale ridge lines) in the Northern Hemisphere is used for making 30-day weather forecasts for Hungary. However, 30-day lead-time is not sufficient plus five-day average data does not allow one to identify the air temperature extremes for the concrete date. In (Kerr 1989), 90-day climate forecasts is used which does not allow to identify heat/cold waves for the concrete date as well. In (Ross 2005), 1 to 4 weeks lead-time is used which is not sufficient for long-term forecast. In general, the analog method is difficult to use because it is impossible to find a perfect analog. Various weather features rarely align themselves in the same locations they were in the previous time. We suggest that our nonanticipative approach, where the current state of a meteorological variable can be described by other related variables with appropriate delay, will generalize and improve the analog methodology.

In (Hennessy et al.2011; Della-Marta and Wanner 2006; Ghil et al. 2011), the extreme value theory is used extensively for identification of climatic extremes at the global and local scales based on the generalized extreme value distribution. Unfortunately, this method of predicting extreme values depends on estimating a probability distribution (which can be time-dependent) from observed values, and even the type of the distribution may be unknown because enough of the right air temperature data is not obtainable. In (Della-Marta and Wanner 2006), a nonlinear model is used for estimation of the relationship between a candidate station and a highly correlated reference station, but the lead-time is only up to 6 days. The present work generalizes the previous method, for application to long-term forecasting, avoiding the requirement for a distribution, and avoiding a specific analytical form for the predictive relationships, which can be too restrictive. According to the present authors, nonanticipative meteorological forecasting systems, as developed in the mathematical forecasting community (e.g. (Greengard and Ruszczynski 2002; Luciano Raso 2013)) now offer the most promise for improving forecast accuracy (Zubov 2013). In principle, it is necessary to identify the form of the dependency between the first meteorological variable's current value (or values if several arguments are taken into consideration) and the future state of the second one (in general, the first and second may be identical) using a "quality criterion" based on repetitions of appropriate patterns (more repetitions within learning sample correspond to higher probability within validation sample). A numerical representation of events is calculated as a sum of the meteorological variables' values. The main problems are based on the formulation of a quality criterion and the selection of input data. Sophisticated methods are needed because enough input data is not obtainable.

In trying to predict the occurrence of extreme events, we are addressing the problem of qualitative forecasting in a particular guise, since such events are likely to be associated with qualitative anomalies in the general circulation. Coherent structures such as blocking patterns play a key role in defining such anomalies. Hence, long-term forecasting of extreme events can be based on the recurrence and temporal patterns of such coherent structures. Where the dynamics is based on wave-propagation in a compact domain, irregular recurrence patterns are indeed to be expected, as with the ENSO cycle. Assumptions of an oscillatory cycle (e.g. Ghil et al. (2011)) may be too restrictive. Statistical analysis, such as that undertaken here, can reveal underlying spatiotemporal structure.

Such statistical analysis has indeed been used to reveal underlying synchronization dynamics (Duane and Tribbia 2004). It has been hypothesized that situations commonly arise where synchronization between two systems can be defined only in terms of coherent structures ("internal synchronization") within each of the synchronizing systems (Duane 2009; Duane 2004). The situation of interest in this paper is one where the "synchronized" events are actually displaced in time. Thus, the statistical approach to long-range forecasting discussed here is believed to be grounded in such subtle features of climate dynamics as also gives rise to the well-known teleconnection patterns and lesser known weak synchronization effects as discussed in (Duane and Tribbia 2004) and (Duane et al. 1999). The displacement in time gives such effects predictive power.

This paper is organized as follows: In Section 2, data sources and principle of nonanticipative analog long-term forecasting of air temperature extremes based on inductive modelling is discussed. In Section 3, the main results and discussion are shown. In Subsection 3.1, nonanticipative long-term forecasting of positive air temperature extremes at Washington National Airport is discussed in detail, and negative extremes are discussed in a more general way. In Subsection 3.2, nonanticipative long-term forecasting of negative air temperature extremes at Skopje Airport is discussed in detail, with a more general discussion of positive extremes. Conclusions are summarized in Section 4.

## 2. Data Sources and Principle of Nonanticipative Analog Long-Term Forecasting of Air Temperature Extremes

2.1 Data sources

NOAA Satellite and Information Service is used as a main data source from 1973 to 2013, providing 66 average daily air temperature datasets from around the world (Zubov 2013), Washington National Airport's mean visibility in miles, mean wind speed in knots, mean dew point in Fahrenheit, maximum and minimum temperatures in Fahrenheit reported during the day, Darwin and Tahiti sea level pressures, southern oscillation index (SOI), equatorial SOI, sea surface temperature, multivariate ENSO index (average monthly). In addition, sea level data (Aburatsu, Japan; http://ilikai.soest.hawaii.edu/woce/wocesta.html; average daily) is used. Hence, 78 datasets are taken into consideration – $X_i=\{x_{i1}, x_{i2}, …, x_{ij}, …\}$, $i = \overline{1,78}$, $j = \overline{1,14975}$ ($j$=1 corresponds to Jan 1, 1973, $j$=14975 – to Dec 31, 2013). These datasets and resources were selected because of free public access and data archived since 1973 at least.

2.2 Principle of Nonanticipative Analog Long-Term Forecasting of Air Temperature Extremes

We assume that some event (or group of events) $A(j)$ has an impact on another event $B(j')$, where $B(j')$ – extreme air temperature (defined as two standard deviations away from the mean), $j$, $j'$ – event dates and $j'$-$j$>0. In our method, four-fold repetitions of extreme air temperature for a give lead time $j'$-$j$ after the same $A(j)$, within the learning sample is taken to establish the validity of $A(j)$ as a predictor. The dependency thus discovered is used as a criterion for prediction. We seek to perform long-term forecasting of Washington National Airport air temperature extremes. Preprocessing standardizes the data using climatological values $\overline{x}_{ij}$ calculated as expectations for the appropriate date:

$$x^*_{ij} = x_{ij} - \overline{x}_{ij}.$$

Climatological values $\overline{x}_{ij}$ are calculated from Jan 1, 1973 to Dec 31, 2013. A value is considered extreme if the difference between this value and its expectation is greater than two standard deviations (SD) in absolute units. This seems a useful definition of extreme events regardless of the form of the distribution. Considering the Washington National Airport dataset, $i$=65. positive $x^+_{65, j_+}$

($j_+ = \overline{1,358}$) and negative $x^-_{65,j_-}$ ($j_- = \overline{1,298}$) extremes are studied, together comprising the set of extremes $E$. ($x^+_{65,j_+}, x^-_{65,j_-} \in E$). Data are split into a learning sample(from 1975 to 2010 yr: $j = \overline{731,13879}$; yrs 1973 and 1974 are reserved because of the lead-time $l$ and the summation interval length $n$ which are up to one year each) and a validation sample(from 2011 to 2013: $j = \overline{13880,14975}$).

Considering the Washington National Airport dataset, the quality criterion that defines an event $A(j)$ as a precursor to an extreme event $B(j')$ is based on situations in the learning sample where the sum of the temperatures at some pair of locations over some time interval prior to some lead-time is unusually large or unusually small, i.e.

$$\sum_{k=0}^{n-1} \left( x^*_{i_1,(j'-k-l)} + x^*_{i_2,(j'-k-l)} \right)\Bigg|_{\substack{i_1,i_2 \in I \\ x^*_{65,j'} \in E \\ j'=[731,13879] \\ l \in L \\ n \in N}} > \underset{i_1,i_2 n,l}{Max}$$

$$\vee \sum_{k=0}^{n-1} \left( x^*_{i_1,(j'-k-l)} + x^*_{i_2,(j'-k-l)} \right)\Bigg|_{\substack{i_1,i_2 \in I \\ x^*_{65,j'} \in E \\ j'=[731,13879] \\ l \in L \\ n \in N}} < \underset{i_1,i_2 n,l}{Min},$$

where

$$\underset{i_1,i_2 n,l}{Max} \equiv \underset{\substack{j=[731,13879] \\ x^*_{65,j} \notin E}}{\max} \sum_{k=0}^{n-1} \left( x^*_{i_1,(j-k-l)} + x^*_{i_2,(j-k-l)} \right)\Bigg|_{\substack{i_1=[1,78] \\ i_2=[i_1,78] \\ l=[14,365] \\ n=[1,365]}},$$

$$\underset{i_1,i_2 n,l}{Min} \equiv \underset{\substack{j=[731,13879] \\ x^*_{65,j} \notin E}}{\min} \sum_{k=0}^{n-1} \left( x^*_{i_1,(j-k-l)} + x^*_{i_2,(j-k-l)} \right)\Bigg|_{\substack{i_1=[1,78] \\ i_2=[i_1,78] \\ l=[14,365] \\ n=[1,365]}}$$

($k$ – temporal summation index (days); $n$ – summation interval length (days); $l = j'-j$ – lead-time (days); $i_1, i_2 \in I$, $n \in N$, $l \in L$ for interrelated sets $I \subset [1,78]$, $N \subset [1,365]$, $L \subset [14,365]$ of meteorological variables, possible lengths of summation intervals, and lead-times respectively). Sums over more than two locations were not considered because of the high computational complexity of the proposed nonanticipative analog algorithm. The sets $I$, $N$, $L$ encompass the precursor events $A(j)$ for a given extreme event $B(j')$, for $j = j' - l$. A given event is defined by a unique tuple ($i_1$, $i_2$, $n$, $l$) together with $Min$ and $Max$. Then, input datasets $i_1$ and $i_2$ with appropriate lead-time $l$ and summation interval of length $n$ are selected to define a prediction rule if the sum of meteorological variables from datasets $i_1$ and $i_2$ is greater than maximum $Max$ (or less than minimum $Min$) four times at least (with a time difference greater than 30 days) within the learning sample, for cases where $x^*_{65,j'} \in E$, i.e. where there is an extreme event $B(j')$ at day $j'$ ($i$=65 is the index of the particular dataset for the forecasting location, Washington). Hence, every extreme selection rule includes six parameters – the indices of two datasets $i_1$ and $i_2$, the lead-time $l$, the summation interval length $n$, maximum *Max*, and minimum *Min*. *Max* and *Min* are computed as the maximum and minimum values of the sums over the datasets $i_1$ and $i_2$, with the same summation interval $n$ and lead-time $l$, where an extreme event does not occur in the learning sample. The gist of the representation in terms of $n$, $l$, $j$, and $j'$ is illustrated in Fig. 1 (calculation of extreme air temperature at Washington National Airport on Apr 16, 2012 using average daily air temperatures from Bucuresti INMH-Bane (Romania, 18) and Kiev (Ukraine, 31) with lead-time 60 days, summation interval 4 days). Note that none of the six parameters are selected arbitrarily; all possible rules are used that satisfy the validation criteria.

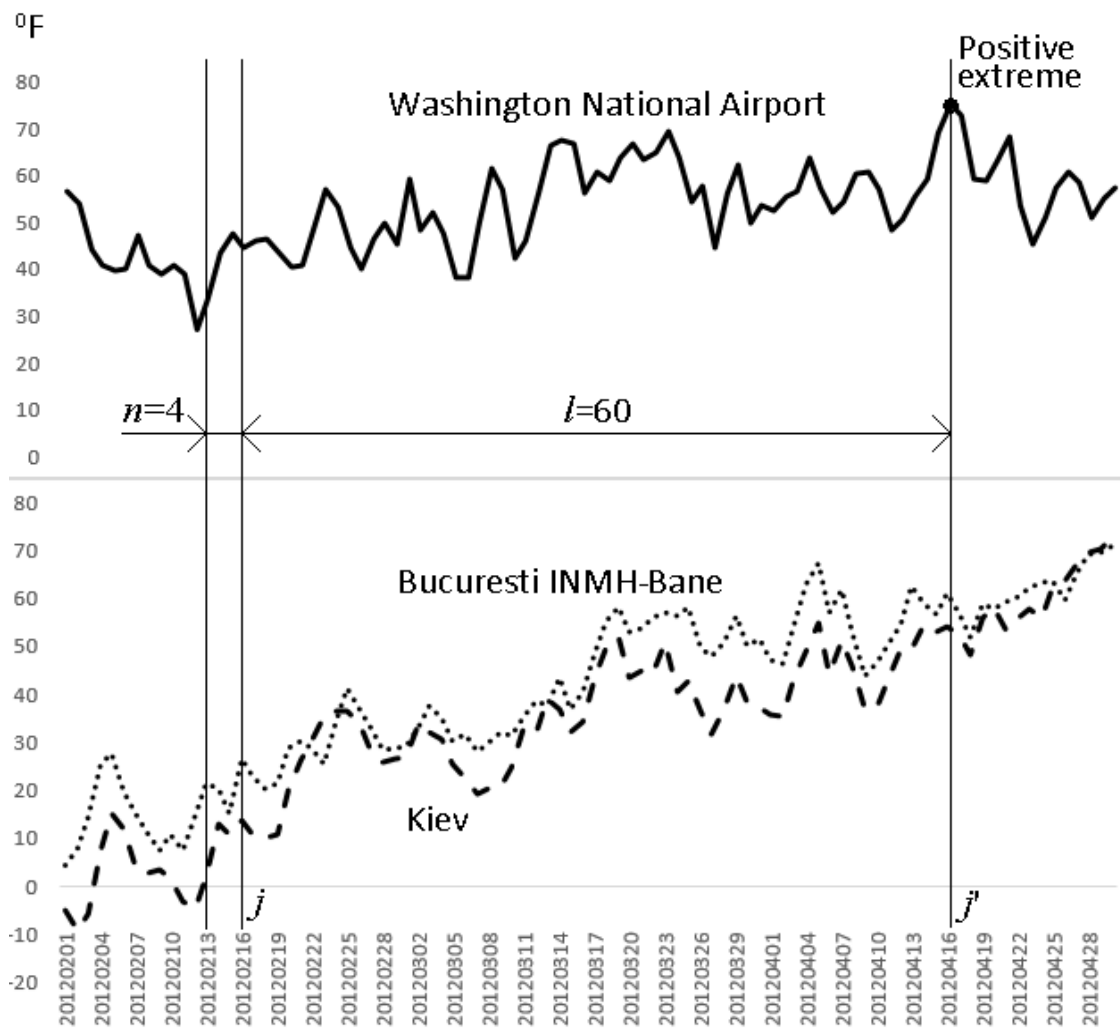

Figure 1. Illustration of variables $n$, $l$, $j$, and $j'$ (calculation of extreme air temperature at Washington National Airport on Apr 16, 2012 using average daily air temperatures from Bucuresti INMH-Bane (Romania, 18) and Kiev (Ukraine, 31) with lead-time 60 days, summation interval 4 days)

## 3. Results and discussion

3.1 Nonanticipative Long-Term Forecast of Washington National Airport Air Temperature Positive Extremes

The three rules for Washington National Airport air temperature positive extremes are:

1. $i_1$=6, $i_2$=57, $n$=1 days, $l$=302 days, $Min$=-34.9$^0$F, $Max$=36.4$^0$F. This rule is described by the tuple (6, 57, 1, 302, -34.9, 36.4) concisely. This rule was in use four times within learning sample on Dec 8, 1978, Nov 26, 1979, Dec 18, 1984, and Oct 1, 1986.

2. (18, 31, 4, 60, -188.1, 138.1). This rule was in use five times within learning sample on Apr 22, 1985, Apr 27-28, 1990, Apr 14, 2002, and Mar 23, 2007.

3. (62, 64, 205, 332, -1619.0, 1741.2). This rule was in use seven times within learning sample on Mar 12-14, 1990, Apr 27, 1990, Apr 9-10, 1991, and Mar 6, 2004.

These rules allow forecasting of positive extremes, as documented in Table 1. All invocations of the above rules are listed in the table. The forecasted extremes were compared to maximal or minimal observed values in the same sector of the month (with months divided into thirds), since prediction of the exact date is not to be expected. Extreme air temperature at Washington National Airport on Apr 16, 2012 was predicted using average daily air temperatures from Bucuresti INMH-Bane (Romania, 18) and Kiev (Ukraine, 31) with lead-time 60 days, summation interval 4 days, minimum sum -188.1$^0$F, maximum sum 138.1$^0$F (see Fig. 1). The same situation was observed five times before – on Apr 22, 1985, Apr 27-28, 1990, Apr 14, 2002, and Mar 23, 2007. Washington National Airport had eleven groups of positive extremes from 2011 to 2013. Hence, the proposed nonanticipative method predicted 18.2 % (including first prediction on Mar 9-31, 2011) of actual positive extremes which is 50 % of all forecasted positive extremes.

The probability of the heat wave occurrence at Washington National Airport is 11.8 % which is calculated as the ratio of the number of heat waves (174) to the number of forecasted sectors from 1973 to 2013 (41*12*3=1476). Here, we assuming that the heat waves, which are typically short, do not span more than one sector. Under a null hypothesis that each invocation of the rules picks a "heat wave" sector randomly, there is an 11.8 % chance that the guess is correct. For the four invocations listed in Table 1, there is a 1-(1-0.118)$^4$=0.395 (39.5 %) chance that at least one forecasted heat wave would be right. The probability of this occurrence, together with a correct prediction of sign for the other three invocations, under the same null hypothesis, is reduced by another factor of 2$^3$=8. The probability of these results with the added condition that one of the other three forecasted sectors is close to extreme is reduced further still. Thus, the probability that random rules give results like those in Table 1 is less than 4.94 %, and our results are significant at the 95 % level.

Table 1. Forecasted Washington National Airport air temperature positive extremes within validation sample (2011-2013 yrs)

| Rule No. ($\overline{1,3}$) | Exact date of extreme value forecast | Forecasted sector of the month | Observed maximum value in forecasted sector of the month, $^0$F | Climatological baseline, $^0$F | SD, $^0$F | Analysis |
|---|---|---|---|---|---|---|
| 3 | Mar 9-31, 2011 | The middle of Mar 2011 | 62.1 (Mar 19, 2011) | 48.0 | 7.6 | Close to extreme value |
| 3 | Apr 1-8, 14-18, 2011 | The middle of Apr 2011 | 62.2 (Apr 4, 2011) | 54.7 | 8.0 | The same sign, difference equals one SD |
| 2 | Apr 2, 3, 11-12, 2012 | The middle of Apr 2012 | 75.0 (Apr 16, 2012) | 57.8 | 8.0 | Extreme value |
| 1 | Dec 2-5, 2012 | The beginning of Dec 2012 | 57.9 (Dec 4, 2012) | 43.8 | 9.0 | The same sign, difference is greater than one SD |

Two rules were identified for Washington National Airport air temperature negative extremes:

1. (35, 59, 16, 287, -161.2, 143.6). This rule was in use five times within learning sample on Aug 29-30, 1986, Mar 15, 1993, Sept 4, 1997, and Oct 30, 2002.

2. (37, 75, 2, 352, -32.4, 41.9). This rule was in use four times within learning sample on Dec 1, 1976, Jan 13, 1977, Jan 28, 1987, and Aug 13, 1996.

In fact, the above two rules did not identify any extreme values within the validation sample. That is probably a result of the small number (298 from 1973 to 2013) of negative extremes.

The nonanticipative long-term forecasting of Washington National Airport air temperature positive extremes has 100% accuracy with respect to the sign of predicted and actual values. The predicted extreme value on Apr 16, 2012 is based on data from proximate times of the year (rule 2: Apr 22, 1985, Apr 27-28, 1990, Apr 14, 2002, and Mar 23, 2007), so seasonal effects are implicitly incorporated.

3.2. Nonanticipative Long-Term Forecast of Skopje Airport Air Temperature Negative Extremes

The five rules for Skopje Airport air temperature negative extremes are as follows:

1. (20, 66, 2, 316, -65.7, 55.9. This rule was in use nine times within learning sample on Dec 10-11, 1983, Jan 10, 1987, Nov 8-9, 1988, and Dec 25-27, and 29, 1998.

2. (20, 66, 3, 315, -96.7, 81.8). This rule was in use nine times within learning sample on Dec 10, 1983, Jan 10, 1987, Nov 7-8, 1988, and Dec 25-29, 1998.

3. (20, 66, 4, 315, -117.7, 99.5). This rule was in use eleven times within learning sample on Dec 10-11, 1983, Jan 10, 1987, Nov 8-9, 1988, and Dec 26-31, 1998.

4. (20, 66, 5, 314, -146.0, 117.2). This rule was in use twelve times within learning sample on Dec 10, 1983, Jan 10, 1987, Nov 7-9, 1988, and Dec 25-31, 1998.

5. (7, 36, 111 days, 274, -500.7, 1273.2). This rule was in use six times within learning sample on Oct 9, 1975, Sept 7, 1976, Apr 28, 1984, and Sept 19-21, 2008.

These rules allow forecasting of negative extremes (see Table 2). Extreme air temperature at Skopje Airport on Dec 15, 2012 was predicted using average daily air temperatures from Busan (South Korea, 20) and Wien-Hohe Warte (Austria, 66) with lead-time 316 days, summation interval 2 days, minimum sum -65.7$^0$F, maximum sum 55.9$^0$F. The same situation was observed nine times before – on Dec 10-11, 1983, Jan 10, 1987, Nov 8-9, 1988, and Dec 25-27, and 29, 1998. In reality, Skopje Airport had nine groups of negative extremes from 2011 to 2013. Hence, the proposed nonanticipative method predicted 11.1 % of actual negative extremes (50 % of all negative forecasted extremes).

Table 2. Forecasted Skopje Airport air temperature negative extremes within validation sample (2011-2013 yrs)

| Rule No. $(\overline{1,5})$ | Exact date of extreme value forecast | Forecasted sector of the month | Observed minimum value in forecasted sector of the month, $^0$F | Climatological baseline, $^0$F | SD, $^0$F | Analysis |
|---|---|---|---|---|---|---|
| 1 | Dec 15, 2012 | The middle of Dec 2012 | 20.6 (Dec 15, 2012) | 34.8 | 6.6 | Extreme value |
| 1, 2 | Dec 13, 2013 | The middle of Dec 2013 | 23.7 (Dec 18, 2013) | 34.9 | 7.5 | The same sign, difference is greater than one SD. Air temperatures' actual values are less than expected for the entire middle of Dec 2013. |
| 1, 3 | Dec 14, 2013 | | | | | |

Probability of the cold wave occurrence at Skopje Airport is 9.8 % which is calculated as the ratio of the number of cold waves (144) to the number of forecasted sectors from 1973 to 2013 (1476).

Similarly to subsection 3.1, the probability 4.23 % that random rules give results like those in Table 2, is less than 5 % and our results are significant at the 95 % level.

Three rules were identified for Skopje Airport air temperature positive extremes:

1. (18, 26, 2, 88, -62.1, 44.8). This rule was in use four times within learning sample on Apr 29, 1983, Apr 11, 1985, May 26, 1990, and Oct 2, 1994.

2. (23, 31, 2, 157, -66.6, 42.5). This rule was in use five times within learning sample on Jun 15-16, 1987, Jun 24, 1993, Jul 11, 2002, and Jun 25, 2007.

3. (42, 44, 98, 181, -826.4, 856.6). This rule was in use five times within learning sample on Sept 17, 1987, Jun 4, 1994, Jul 23, 2007, and Dec 1 and 24, 2010.

In fact, the above three rules did not identify any extreme values within the validation sample. That is probably a result of the small number (238 from 1973 to 2013) of positive extremes.

The nonanticipative long-term forecast of Skopje Airport air temperature negative extremes has 100% accuracy with respect to the sign of predicted and actual values. Again the rule used to predict the predicted extreme value on Dec 15, 2012 is based on data from proximate times of the year (rule 1: Dec 10-11, 1983, Jan 10, 1987, Nov 8-9, 1988, and Dec 25-27, 29, 1998).

## 4. Conclusions

In this paper, a nonanticipative analog forecasting system for air temperature extremes has been proposed for the improvement of forecast accuracy toward mitigating the very deleterious socioeconomic effects of those extremes. The method is based on identification of dependencies between the current value(s) of one or more meteorological variables (here two variables) and the future state of another variable (which may be identical to the first, as with air temperatures used here). The main issues are the formulation of a quality criterion and selection of input data. The method was applied to the prediction of positive extremes for Washington National Airport and negative extremes for Skopje Airport. The data included standard meteorological variables from 66 places around the world, as well as sea level (Aburatsu, Japan), average monthly Darwin and Tahiti sea level pressures, SOI, equatorial SOI, sea surface temperature, and multivariate ENSO index.

The data is split into two samples, for learning and validation, respectively. Initially, the sum of the values at two different locations (minus corresponding expectation values) is calculated with lead-time from 14 to 365 days, and summed over an interval from 1 to 365 days. The criterion selects datasets from two locations with appropriate lead-time and summation interval, which have more than the maximum (or less than the minimum) sum over the rest of data four times at least (with a minimum time difference of at least 30 days), when a later extreme event occurs in the learning sample, thus defining rules that are applied to the validation sample. Specific extreme events at Washington National Airport and at Skopje Airport were predicted using the rules thus defined.

Similar results, not reported here, were achieved for Kiev (Ukraine). Some extremes are specifically predicted (about 10 % of all extremes). The methodology has 100 % forecast accuracy with respect to the sign of predicted and actual values. That is, for all "false alarms" (about 50 % – as specifically listed in Tables 1 and 2), at least the sign is correct. Surely, one would gain more by preparing for the correctly predicted extremes than one would lose by taking unnecessary measures for the same number of false alarms. No parameters in the rules, or in the procedure used to derive them, are chosen arbitrarily.

The method might be further validated using simulated data, a possibility we intend to explore in future studies. However, the relevance of simulated data for the study of extreme events is questionable because of the high degree of nonlinearity of the processes involved. We expect that models may not capture essential elements of these processes because enough of the data that would be needed for the construction of predictive models is fundamentally not obtainable.

The most likely prospect for application of this work is the development of a global scale nonanticipative long-term forecasting system for air temperature extremes. It is expected that geographical and temporal patterns linking observed and future events through appropriately defined quality criteria will be found to extend and improve the simple algorithm described here.

**Acknowledgements**

This work was jointly supported by the Microsoft Azure for Research program (Climate Data Initiative Award) and the University of Information Science and Technology "St. Paul the Apostle".